\documentclass[fleqn,usenatbib]{mnras}

\usepackage{newtxtext,newtxmath}
\usepackage[T1]{fontenc}

\DeclareRobustCommand{\VAN}[3]{#2}
\let\VANthebibliography\thebibliography
\def\thebibliography{\DeclareRobustCommand{\VAN}[3]{##3}\VANthebibliography}

\usepackage{graphicx}	
\usepackage{amsmath}	
\usepackage{bm}
\usepackage{hyperref}
\usepackage{graphicx}
\usepackage{xspace}
\usepackage{url}
\usepackage{cancel}
\usepackage{changes}
\usepackage{soul}

\newcommand{\hatz}{ \widehat{z} }
\newcommand{\hatrsh}{ r_{\rm sh,250} }
\newcommand{\gm}{ \gamma_{\rm e,min} }
\newcommand{\gM}{ \gamma_{\rm e,max} }

\newcommand{\Msun}{\ {\rm M}_\odot }
\newcommand{\mpy}{\ \Msun\, {\rm yr}^{-1} }
\newcommand{\kmps}{\ { \rm km \, s}^{-1} }
\newcommand{\pcMpc}{\ {\rm Mpc}^{-3}}

\newcommand{\ergpcc}{\mbox{ erg cm}^{-3} }
\newcommand{\beqr}{\begin{eqnarray} \nonumber}
\newcommand{\eeqr}{\end{eqnarray}}
\def\beq{\begin{equation}}
\def\eeq{\end{equation}}
\def\beqn{\begin{eqnarray}}
\def\eeqn{\end{eqnarray}}

\usepackage{ulem}
\newcommand{\sy}[1]{\textcolor{red}{\bf{#1}}}

\title[ORCs from virial shocks?]{Are Odd Radio Circles virial shocks around massive galaxies?\\ Implications for cosmic-ray diffusion in the circumgalactic medium
}

\author[Yamasaki, Sarkar \& Li]
{Shotaro Yamasaki\thanks{E-mail: shotaro.s.yamasaki@gmail.com}$^{1}$, Kartick C. Sarkar$^{2,3,4}$ and Zhaozhou Li$^{3}$
\\
$^{1}$Department of Physics, National Chung Hsing University, 145 Xingda Rd., South Dist., Taichung 40227, Taiwan\\
$^{2}$School of Physics and Astronomy, Tel Aviv University, Tel Aviv, 6997801, Israel\\
$^{3}$Racah Institute of Physics, The Hebrew University of Jerusalem, Jerusalem 91904, Israel\\
$^4$Dept. of Space, Planetary \& Astronomical Sciences and Engineering, Indian Institute of Technology Kanpur, 208016, India
}

\date{Accepted XXX. Received YYY; in original form ZZZ}

\pubyear{2023}

\begin{document}
\label{firstpage}
\pagerange{\pageref{firstpage}--\pageref{lastpage}}
\maketitle

\begin{abstract}
Recently, a new population of circular radio ($\sim$GHz) objects have been discovered at high Galactic latitudes, called the Odd Radio Circles (ORCs). A fraction of the ORCs encircles massive galaxies in the sky with stellar mass $\sim 10^{11}\Msun$ situated at $z=0.2$--$0.6$, suggesting a possible physical connection. 
In this paper, we explore the possibility that these radio circles originate from the accretion shocks/virial shocks around massive ($\gtrsim10^{13}\,\Msun$) dark matter halo at $z\sim0.5$. 
We found that the radio flux density of the emitting shell is marginally consistent with the ORCs. We also find that pure advection of electrons from the shock results in a radio-emitting shell that is considerably narrower than the observed one due to strong inverse-Compton cooling of electrons. Instead, we show that the diffusion of cosmic-ray (CR) electrons plays a significant role in increasing the width of the shell. We infer a diffusion coefficient, $D_{\rm cr} \sim 10^{30}\ {\rm cm^2\,s^{-1}}$, consistent with the values expected for low-density circumgalactic medium (CGM). If ORCs indeed trace virial shocks, then our derived CR diffusion coefficient represents one of the few estimations available for the low-density CGM.
Finally, we show that the apparent discrepancy between ORC and halo number density can be mitigated by considering an incomplete halo virialization and the limited radiation efficiency of shocks. This study, therefore, opens up new avenues to study such shocks and non-thermal particle acceleration within them. Furthermore, our results suggest that low-mass galaxies ($\lesssim 10^{13}\Msun$) may not show ORCs due to their significantly lower radio surface brightness.

\end{abstract}




\section{Introduction}
Recently, mysterious diffuse radio circles,
called Odd Radio Circles (ORCs), have been discovered by radio surveys \citep{norris21a,norris21b,koribalski21,norris22,filipovic22,omar22_LOFAR,koribalski23}.
ORCs have sizes $\sim$ arcminutes and have radio brightness $\sim 2$--$9$ mJy at GHz frequencies.
They are mostly found at high Galactic latitudes with no association with any known sources. Recent observations, however, have discovered massive galaxies (stellar mass, $M_\star \sim 10^{11} \Msun$ and redshift, $z=0.2$--$0.6$) at the center of some of the ORCs \citep{norris22}, thus fueling the idea that ORCs may be related to the massive galaxies.
If associated with the central galaxies, their approximately circular and edge-brightened structure would have a ring radius of $\sim 200$ kpc.

Although there is a lack of detected features in other wavebands, three out of seven ORCs are known to contain galaxies at their geometrical center in the sky, suggesting a possibility of their extragalactic origin. 
One of these sources, ORC J2103-6200 (hereafter ORC1; \citealt{norris21a,norris22}), has been most intensively observed both in radio and optical bands. It is found to possess a non-thermal radio spectrum, with an observed spectral index, $\alpha$, varying from $-1.5$ to $-1.3$ (where flux density $F_\nu\propto\nu^\alpha$) over the frequency range of $0.1$--$2$ GHz. 
Polarization measurements revealed the existence of tangential (to radial direction) magnetic fields, implying the existence of a strong shock. The apparent thickness of the radio-emitting shell is estimated to be about $10$--$20\%$ of the radius (after deconvolving the effect of the antenna beam size).

Theoretical interpretations of ORCs are divided depending on whether their origins are assumed to be local (i.e. within the Galaxy or its immediate neighborhood) or extragalactic. The former includes supernova remnants from the Local Group \citep{filipovic22,omar22snr,sarbadhicary22}, whereas the latter includes the forward/termination shock driven by an old starburst event \citep{norris21a,norris22} and by transient events such as binary supermassive black hole (BH) mergers \citep{koribalski21,norris22}, multiple tidal disruptions of stars by an intermediate-mass BH \citep{omar22tde} and galaxy mergers \citep{dolag22}.

One of the theoretical models within the extragalactic scenario is the star formation-driven forward/termination shock from an old 
starburst event with a star formation rate, SFR $\sim 100 \mpy$ in a massive galaxy ($M_\star \approx 3\times 10^{11} \Msun$) \citep{norris22}. 
However, such a massive galaxy is expected to contain a hot circumgalactic medium (CGM) with sound speed, $c_{\rm s} \sim 300 \kmps$, and the forward shock would fade away\footnote {As the shock expands in an ambient medium with density profile shallower than $r^{-2}$, the shock speed reduces with time and after a certain time the shock speed becomes comparable to the sound speed of the medium. At this point, the shock can no longer be distinguished from the ambient medium. This time is called the fade-away time \citep{Draine2011, Dekel2019}} in $\sim 500$ Myr \citep{Sarkar2015, Lochhaas2018}.  
Assuming that the cosmic-ray (CR) particles were accelerated during this time, the synchrotron signatures of such shocks would be visible till the cooling time of the CR population. Since any shocks with $\sim\mu$G magnetic field strength, which is typical for such shocks, cools over $\sim 100$ Myr (see equation \ref{eq:t_cool}), the synchrotron signatures of such energetic events would not last for more than $\sim 600$ Myr. 
Therefore, the scenario is inconsistent with a very old ($\sim 5$ Gyr) starburst.

Recently, \citet{dolag22} proposed a possibility of merger-driven internal shocks as an origin of ORC structures with $M_{\rm vir}=10^{12}\,\Msun$. While it successfully accounts for both the rarity of ORCs and the complex inner sub-structure seen in ORCs, they found much fainter radio rings than observed. Here, we explore an alternative possibility within a parallel conceptual framework centered on galactic-scale shocks. We posit that ORCs might represent large-scale (radius of $\sim 200$ kpc) accretion/virial shocks around massive galaxies. 
In this scenario, the observed radio emission is due to the synchrotron emission from CR electrons accelerated at the virial shock around the galaxy.

The diffuse non-thermal emission from virial shocks has been intensively discussed in the literature on galaxy clusters. In such a scenario, highly relativistic electrons accelerated at the intergalactic shock cool either via inverse-Compton scattering with the Cosmic Microwave Background (CMB) photons and/or the synchrotron radiation. The former could be observable as a high-energy gamma-ray background \citep{loeb00,totani00,keshet03,keshet17,keshet18} and the latter as the radio background \citep{waxman00,keshet04} or as extended radio sources \citep{hoeft07,marinacci18}. 
While there is tentative observational evidence for the virial shock emission from clusters with halo mass of $\sim 10^{15}\Msun$ \citep[e.g.,][]{keshet17,keshet18}, it has never been clearly detected so far from galaxies that are far less massive. Therefore, if our scenario is the case for at least some fraction of the ORCs, it would be the first direct evidence of virial shocks in less massive haloes.

This paper is organized as follows. We present our dynamical and emission models in \S \ref{s:emission} and the number density of ORCs predicted by our model in \S \ref{s:density}.  
We summarize our findings with discussion in \S \ref{s:discussion}. Throughout this work, we assume a $\Lambda$CDM cosmology with $\Omega_{\rm m}=0.3$, 
$\Omega_{\Lambda}=0.7$, and $h=H_0/(100 \; {\rm km\,s^{-1}\,Mpc^{-1}})=0.7$, and cosmological baryon fraction $f_{\rm b}\equiv\Omega_{\rm b}/\Omega_{\rm m}\sim0.16$ for simplicity. 

\begin{table*}
\centering
\caption{A summary of the properties of the four published ORCs that contain central galaxies with their photometric redshifts. The parameters above and below the double lines are observed and estimated quantities, respectively. Halo masses are estimated by comparing $r_{\rm sh}=f_{\rm sh}\,r_{\rm vir}$ (see equation \ref{eq:r_sh}) with observed physical radii of ORCs. Electron spectral index $s$ is determined assuming that the observing frequency lies above the cooling frequency (equation \ref{eq:nu_c}) where $\alpha=-s/2$. References: [1] \citet{norris21a}; [2] \citet{norris22}; [3] \citet{koribalski21}.
}
\setlength{\tabcolsep}{3pt} 
\renewcommand{\arraystretch}{1.25} 
\label{tab:summary}
\begin{tabular}{lccc}
\hline
Source name  & ORC J2103-6200 (ORC1)  & ORC J1656+2726 (ORC4)    & ORC J0102-2450 (ORC5) 
\\   
\hline\hline
\multicolumn{4}{c}{ORC properties}
\\
\hline
Angular radius (arcsec)                          & 40                & 45           & 35         
\\
Flux density (mJy at 1 GHz)                 & 3.9                & 9.4                & 3.2  
\\
Spectral slope $\alpha$ ($S_\nu\propto\nu^{\alpha}$) & $-1.4\pm0.1$       & $-0.9\pm0.2$       & $-0.8\pm0.2$     
\\
Observation frequency range $\nu_{\rm obs}$ (GHz) & $0.1$--$2$       & $0.8$--$1.1$       & $0.8$--$1.1$  
\\
\hline
\multicolumn{4}{c}{Host (central) galaxy candidate properties}
\\
\hline
Redshift $z$                                   & 0.55               & 0.39               & 0.27 
\\
Galaxy stellar mass $M_\star$ ($\Msun$)               & $3\times10^{11}$   & ?                   & $1\times10^{11}$ 
\\
Refs.    & [1, 2]   &    [1, 2]              & [2, 3]   
\\\hline\hline
\multicolumn{4}{c}{Inferred quantities from  $z$ and $\alpha$}
\\
\hline
Physical radius (kpc)                          & 260                & 240           & 140  
\\
Galaxy halo mass $ M_{\rm vir}$ ($\Msun$)        & $3.7\times10^{12}\,f_{\rm sh}^{-3}$ & $2.5\times10^{12}\,f_{\rm sh}^{-3}$ & $4.9\times10^{11}\,f_{\rm sh}^{-3}$ 
\\

Electron index $s$ ($dn_{\rm e}/d\gamma_{\rm e}\propto\gamma_{\rm e}^{-s}$)               & $2.8\pm0.2$        & $1.8\pm0.4$        & $1.6\pm0.4$   \\
\hline
\end{tabular}
\end{table*}

\section{Emission from virial shocks}
\label{s:emission}

\subsection{Characteristics of virial shocks}
\label{subsec:virial-shocks}

The virial radius, $r_{\rm vir}$, of a dark matter halo of mass, $M_{\rm vir}$, is defined as
the radius within which the average matter density, $\overline{\rho}_{\rm dm}$, becomes $\approx200$ times\footnote{While the exact value of 200 is subject to variation in different works due to cosmological considerations, the adjustment of this number is accommodated by the inclusion of the factor in $f_{\rm sh}$. Therefore, any alteration in the virial definition would essentially manifest as a modification in the $f_{\rm sh}$ parameter, without substantially impacting the conclusions drawn in this study.} larger than the critical density, $\rho_{\rm c}$, of the universe i.e., $\overline{\rho}_{\rm dm}=M_{\rm vir}/(4/3\pi r_{\rm vir}^3)={200}\,\rho_{\rm c}$. Here, $\rho_c(z)=3 H_0^2\left[\Omega_{\rm m}(1+z)^3+\Omega_{\Lambda}\right]
/(8\pi G)$ with $G$ being the gravitational constant. Hereafter, we use $\Omega_{\rm m}(1+z)^3+\Omega_{\Lambda}\approx (1+z)^{3/2}$ for analytical convenience, which holds to within $\lesssim9$\% accuracy at $z<1$. With these definitions, the halo virial mass
and virial radius are related via
\beqn
\label{eq:r_sh}
r_{\rm vir}\sim410\ {\rm kpc}\ \ M_{\rm vir,13}^{1/3}\,  h_{70}^{-1}\, \hatz^{-1/2}\ ,
\eeqn
where $h_{70}=h/0.7$, $\hatz=(1+z)/1.5$, and $M_{\rm vir,13}=M_{\rm vir}/(10^{13}\Msun \,h^{-1})$ is the halo mass, which is implied by the observed stellar mass of the central galaxies in ORCs $M_\star\sim10^{11}\Msun$ (see \S \ref{s:discussion} for inferred $M_{\rm vir}$ for ORCs). 
The virial shock is assumed to be an accretion shock around the galaxy and is created by continuous baryonic mass accretion onto the galaxy. The size of this shock increases slowly with time but practically remains constant over $\sim$ Gyr time scales for massive galaxies at $z\lesssim 1$\citep{Birnboim2003, dekel06}. Hydrodynamical simulations suggest that the actual radius of accretion shock, $r_{\rm sh}$, could deviate from the virial radius depending on the redshift, mass, and the presence of radiative cooling \citep{keshet04,dekel06,wise07}\footnote{\citet{keshet04} estimate the typical range the shock geometry parameter as $f_{\rm sh}=r_{\rm sh}/r_{\rm vir}\sim0.6$--$1.2$ based on
the simulation of cluster ($M_{\rm vir}\sim10^{\rm 15}\,\Msun$) accretion
shocks \citep{keshet03}. This range of values should also be compatible with $M_{\rm vir} \sim 10^{13} \Msun$ haloes \citep{Birnboim2003}.} and feedback \citep{fielding2017}. The above picture is broadly consistent with the observationally inferred physical radii of ORCs ($140$--$260$ kpc), which are smaller than $r_{\rm vir}$ for $10^{13}\Msun$ haloes. Hereafter, we assume that $r_{\rm sh}$ approximately represents the ORC ring radius for simplicity.

The observed quantities for three ORCs reported with central galaxies are summarized in Table \ref{tab:summary}. 
Assuming that the central galaxies of ORCs are indeed their host galaxies, their physical radii are estimated to be around a few hundred kpc at $z=0.3$--$0.6$. If we interpret this as the shock radius due to accretion, the halo mass inferred via equation \eqref{eq:r_sh} would be $M_{\rm vir}=(0.5$--$4)\times10^{12}f_{\rm sh}^{-3}\Msun$, where $f_{\rm sh}\equiv r_{\rm sh}/r_{\rm vir}$ is the shock geometry parameter. As $f_{\rm sh}\lesssim1$ in general, the actual halo mass could be as high as $10^{13}\Msun$. Given the large dispersion in the stellar-to-halo mass relation, i.e. $M_\star/M_{\rm vir}$,  at these masses \citep[e.g.,][]{moster10,wechsler2018,girelli20}, the inferred halo mass
is broadly consistent with the observed stellar mass of ORC host galaxies $\sim10^{11}\Msun$ (except for ORC4's central galaxy whose stellar mass is unknown). 

As mentioned earlier, the accretion shock practically remains stationary in the galaxy frame. Therefore, the shock speed in the frame of the infalling material is the same as the speed of the infalling material in the galaxy frame. This speed is close to the circular speed of the halo i.e.,
\beqn
\label{eq:v_sh}
v_{\rm sh} \approx \sqrt{\frac{G M_{\rm dm}(<r_{\rm sh})}{r_{\rm sh}}}  \sim 420\ {\rm km\ s^{-1}}\ M_{\rm vir,13}^{5/12}\ \hatrsh^{-3/20}\,h_{70}^{-1/4}\, \hatz^{1/8}.
\eeqn
Here, $M_{\rm dm}(<r_{\rm sh})$ is the enclosed mass inside $r_{\rm sh}$ and we used $M_{\rm dm}(<r_{\rm sh})/M_{\rm vir}=(r_{\rm sh}/r_{\rm vir})^{0.7}$ based on an approximation of Navarro–Frenk–White (NFW) mass profile (\citealt{nfw97,klypin01}; see also Appendix \ref{s:scalings}), and $\hatrsh\equiv r_{\rm sh}/(250\,{\rm kpc})$. 
Note that the speed of the downstream material in the galaxy frame is only $v_{\rm d}=v_{\rm sh}/4$, following the Rankine-Hugoniot jump condition.
Equation \ref{eq:v_sh} clearly indicates that the virial shock is non-relativistic and that the dynamical time for the shocked material to propagate from the shock front to the galactic center only by advection is
\beqn
\label{eq:t_dyn}
t_{\rm dyn}\approx\frac{r_{\rm sh}}{v_{\rm d}}
\sim 2.3\  {\rm Gyr}\ \ M_{\rm vir,13}^{-5/12}\, \hatrsh^{23/20}\,\,h_{70}^{1/4}\,\hatz^{-1/8}\ .
\eeqn
As the infalling gas passes through the virial shock, about $3/4$ of its kinetic energy is converted into thermal energy in the post-shock gas: $u_{\rm th}=(9/32)\,\rho_{\rm sh}v_{\rm sh}^2\sim3.9\times10^{-13}\  {\rm erg\,s^{-1}\ }M_{\rm vir,13}^{4/3}\,\hatrsh^{-9/5}\,  \hatz$, 
where $\rho_{\rm sh}\approx 0.7\,f_{\rm b}\,\overline{\rho}_{\rm dm}\,(r_{\rm sh}/r_{\rm vir})^{-1.5}$ (see Appendix \ref{s:rho_sh}) is the downstream matter density (i.e., the gas density of the CGM in hydrostatic equilibrium with NFW dark matter profile) just behind the shock, and we assume that the upstream speed in the post-shock frame is $(3/4)v_{\rm sh}$, so the mean post-shock energy per particle is $(9/16)m_{\rm p}v_{\rm sh}^2/2$.
Now, as the upstream material crosses through the virial shock, some of its kinetic energy is also converted into magnetic energy and CR energy. It is difficult to predict how much of the shock energy is converted into non-thermal energies as the conversion depends on several uncertain plasma processes and CR acceleration efficiencies. For simplicity, we assume that the magnetic field energy density in the downstream region is only a fixed fraction, $\xi_{B}$, of the thermal energy density i.e., $u_B = B^2/(8\pi)=\xi_{\rm B}\,u_{\rm th}$. This implies
\beqn
\label{eq:B-field}
B=\sqrt{(9/4)\pi \,\xi_{\rm B}\,\rho_{\rm sh}\,v_{\rm sh}^2}
\sim 1 \ {\rm \mu G}\ \ M_{\rm vir,13}^{2/3}\,\hatrsh^{-9/10}\, \,\xi_{\rm B,-1}^{1/2}\,\hatz^{1/2},
\eeqn
where $\xi_{\rm B,-1}=\xi_{\rm B}/0.1$. Similarly, we assume that the energy density of the CR electrons, $u_e$, in the post-shock region is also a fraction, $\xi_e$, of the thermal energy density, $u_{\rm th}$, i.e. $u_{\rm e} = \xi_e\: u_{\rm th}$.

Ideally, we require the expression for the thermal energy density to change since some of the shock energy is now transferred to the magnetic and CR energy densities. However, it can be easily shown for strong shocks that the post-shock thermal energy density increases by a factor of $\approx \mathrm{Mach}^2$, whereas, the magnetic energy density increases by a factor of $16$ \citep{Draine2011} due to a factor of 4 increase in density.  For our virial shocks, Mach $\sim 10$ (considering a pre-shock temperature of $\sim 10^5$ K). This means that the post-shocked gas in our cases will be \sy{\bf{weakly magnetic}} and hence the correction for the magnetic energy density to the thermal energy density (particularly in $u_{\rm th}$) can be ignored. For the CRs, observational evidence puts CR energy density to be about $\sim 10\%$ of the thermal energy. Therefore, a similar correction for the CR energy density can also be ignored. 

\subsection{Emission processes}
\label{subsec:emission-processes}

The synchrotron emission for an electron with Lorentz factor (LF), $\gamma_{\rm e}$, peaks at the characteristic frequency in the observer frame $\nu_{\rm syn}(\gamma_{\rm e}) \approx \gamma_{\rm e}^2 e B/(2\pi m_e c)/(1+z)$
, where $e$ is the electron charge, $m_{\rm e}$ is the electron mass, and $c$ is the speed of light.
The required LF for the electron to produce the observed emission at frequency $\nu_{\rm obs}$
is (using equation \ref{eq:B-field})
\beqn
\label{eq:gamma_obs}
\gamma_{\rm e,obs}
\sim2.3\times10^4\ \ \xi_{\rm B,-1}^{-1/4}\,\nu_{9}^{1/2}\,\hatrsh^{9/20}\,\hatz^{1/4}\ , 
\eeqn
where $\nu_9=\nu_{\rm obs}/10^9$ Hz. This implies that virial shocks if being observed as ORCs, must have an ultra-relativistic population of CR electrons. Since the speed of the upstream material ($\sim 450 \kmps$; see equation \ref{eq:v_sh}) is far more than the typical sound speed of the material ($c_{\rm s}\sim 50 \kmps$, corresponding to a temperature of $10^5$ K), the accretion shock is expected to be strong (Mach $\sim 10$) and hence is an ideal location where CR particles can be accelerated. 
While the above is true for a single electron, a proper calculation of the synchrotron emission requires the knowledge of the electron population.

As the freshly accelerated CR electrons flow with the downstream material, the electrons cool via Inverse Compton (IC) scattering with background CMB photons and synchrotron radiation at a rate
\beqn
\label{eq:C_cool}
\dot{\gamma}_{\rm e}=-\frac{4\sigma_{\rm T}}{3m_{\rm e}c}(u_{\rm CMB}+u_{\rm B}) \,\gamma_{\rm e}^2\approx -\frac{4\sigma_{\rm T}}{3m_{\rm e}c}u_{\rm CMB} \,\gamma_{\rm e}^2\ ,  
\eeqn
where $u_{\rm CMB}=2.19\times10^{-12}\: \hatz^4\, \ergpcc$
is the CMB energy density, and $\sigma_T$ is the Thompson scattering cross-section.
It is clear that $u_{\rm CMB} \gg u_{\rm B}$ for typical parameters of our interest. Therefore, the total cooling of the CR electrons is dominated by the IC process and we can safely neglect the synchrotron cooling in the analytic expression hereafter (but included in the numerical computation).
The cooling time of electrons that produce observed synchrotron emission is 
\beqn
\label{eq:t_cool}
t_{\rm IC}(\gamma_{\rm e}) \sim \frac{\gamma_{\rm e}}{|\dot{\gamma}_{\rm e}|}
\sim 39 \ {\rm Myr}\  \ \gamma_{{\rm e},4}^{-1}\,\hatz^{-4}\ .
\eeqn
Comparing the cooling time of electrons $t_{\rm IC}(\gamma_{\rm e})$ with the dynamical timescale (equation \ref{eq:t_dyn}), one obtains the characteristic cooling LF, $\gamma_{\rm e, cool}\sim1.9\times10^2\ \ M_{\rm vir,13}^{5/12}\,\hatrsh^{-23/20}\,h_{70}^{-1/4}\,\hatz^{-31/8}$. 
Therefore, only those electrons with $\gamma_{\rm e}>\gamma_{\rm e,cool}$ can cool, for which the observed synchrotron emission is expected to be above the cooling frequency
\beqn
\label{eq:nu_c}
\nu_{\rm c} &\equiv& \nu_{\rm syn}(\gamma_{\rm e, cool})\\
&\sim&67\ {\rm kHz}\ \ M_{\rm vir,13}^{3/2}\, \xi_{\rm B,-1}^{1/2}\,\hatrsh^{-16/5}\,h_{70}^{-1/2}\,\hatz^{-33/4}\ . \nonumber
\eeqn
Namely, the observing radio frequencies at GHz are much above the break frequency.  This means that we must account for the cooling of the CRe population.

Let us assume that the freshly accelerated CR electron population within the downstream material between LF $\gamma_{\rm e}$ and $\gamma_{\rm e}+d\gamma_{\rm e}$ can be described by a power-law distribution with a spectral index of $s$ ($2 < s < 3$)\footnote{As shown in Table \ref{tab:summary}, the observing radio frequency is much greater than the cooling frequency at $\nu>\nu_{\rm cool}$ (see equation \ref{eq:nu_c}), where the synchrotron spectrum has a slope of $\alpha=-s/2$ ($F_{\nu}\propto \nu^{\alpha}$). This simple interpretation implies an electron index of $s\sim1.6$--$2.8$ for three ORC sources. Particularly, ORC4 and ORC5 show an electron index $s\lesssim 2$, which might be too hard to produce by diffusive shock acceleration (DSA), the mechanism thought to be responsible for most synchrotron sources. However, given the large uncertainties ($\pm0.2$) in $\alpha$ and in $s$ ($\pm0.4$) measurements, it could be marginally consistent with $s\simeq2$.}, given by
\beqn
n_{\rm e}(\gamma_{\rm e},t=0)=n_{\rm e0}\,\gamma_{\rm e}^{-s}\quad {\rm (\gm<\gamma_{\rm e})}\ , 
\eeqn
where $\gm$ and $t$ are the minimum (typical) electron LF and the proper time measured since the acceleration at the shock front, respectively.  
The electron number density is normalized by
\beqn
\label{eq:n_e0}
\int_{\gm}^{\gamma_{\rm e,max}} d\gamma_{\rm e}\, {d n_{\rm e} \over d\gamma_{\rm e}}\gamma_{\rm e} \,m_{\rm e} c^2  =u_{\rm e} = \xi_{\rm e}\,u_{\rm th} \ .
\eeqn
We note that $\gm$ ($<\gamma_{\rm e,obs}$) is one of the most critical parameters to determine the resulting radio flux through the above normalization and we treat it as a free parameter, unlike the common assumption of $\gm\approx1$, often made for non-relativistic shocks. 
The maximum energy of accelerated electrons is determined by the fact that above a certain energy, the electrons may cool faster than they are accelerated. This maximum energy is estimated by equating $t_{\rm IC}(\gamma_{\rm e})$ (equation \ref{eq:t_cool}) and the CR acceleration timescale $t_{\rm acc}(\gamma_{\rm e})={6\,m_{\rm e}c^3\gamma_{\rm e}/(e B v_{\rm sh}^2})$ \citep[e.g.,][]{totani00, waxman00}: $\gM \sim 9.5\times10^6\ \ M_{\rm vir,13}^{3/4}\,\xi_{\rm B,-1}^{1/4}\,\widehat{r}_{\rm sh}^{-3/4}\,h_{70}^{-1/4}\,\widehat{z}^{-13/8}$. We can effectively consider $\gM$ to be infinitely large as discussed in Appendix \ref{s:emissivity}.

\begin{figure}
\centering
\includegraphics[width=0.52\textwidth, clip=true, trim={5.5cm 4cm 5cm 2.5cm}]{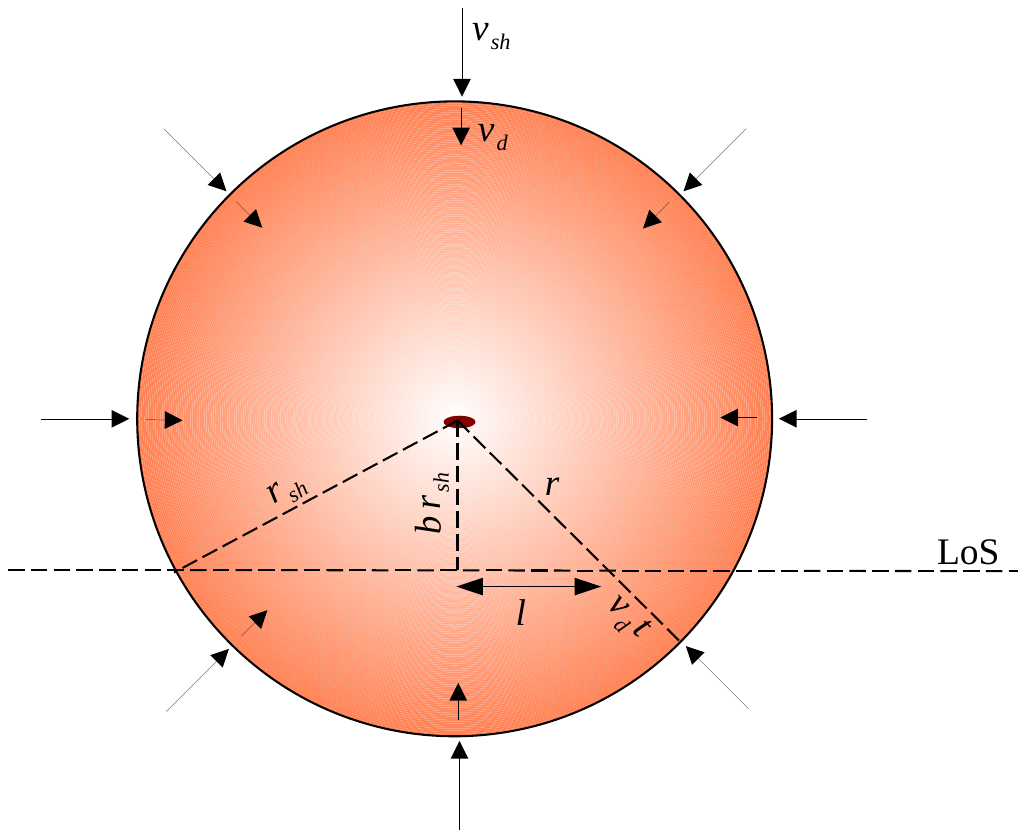}
\caption{Geometry of the accretion shock and line of sight (LoS) integration. Along LOS with impact parameter $br_{\rm sh}$, the distance from the galactic center at a given time $t$ is denoted as $r$ (see equation \ref{eq:advection length}). 
}
\label{fig:geometry}
\end{figure}

We use an analytic formalism to solve for the evolution of the electron spectrum. 
Neglecting the adiabatic cooling/heating and using equation \eqref{eq:C_cool} for cooling term $\dot{\gamma}_{\rm e}$, the time-evolution of electron spectrum at $\gamma_{\rm e, min}<\gamma_{\rm e}$ is given by \citep[][]{Kardashev62,sarazin99}
\beqn
\label{eq:n_e_primary}
n_{\rm e}(\gamma_{\rm e},t)=\begin{cases}
n_{\rm e0}\,\gamma_{\rm e}^{-s}\left(1- b_{\rm IC}\gamma_{\rm e}t\right)^{s-2} &\left(b_{\rm IC}\gamma_{\rm e}t<1\right), \\
0 &\left(b_{\rm IC}\gamma_{\rm e}t>1\right).
\end{cases}
\eeqn

The angle-averaged (assuming isotropic pitch angle distribution) synchrotron power of a single electron with LF $\gamma_{\rm e}$ at observing frequency $\nu=\nu^{\prime}/(1+z)$ is given by 
\citep[e.g.,][]{rybicki79}
\beqn
\label{eq:P_syn}
P_{\nu^{\prime}}^{\prime}(\gamma_{\rm e})=\frac{2\sqrt{3} e^3 B}{3\,m_{\rm e} c^2}F\left(\frac{\nu^\prime}{\nu_{\rm syn}^\prime}\right)\equiv C_{\rm syn}F\left(\frac{\nu }{\nu_{\rm syn}}\right),
\eeqn
where $F(x)=x\int_x^\infty d\xi K_{5/3}(\xi)$ is the synchrotron function with $K$ being the modified Bessel function, and $\nu_{\rm syn}^\prime=(1+z)\nu_{\rm syn}$.
Combining equations \eqref{eq:n_e_primary} and \eqref{eq:P_syn}, the specific synchrotron emissivity can be recast as
\beqn
j^\prime_{\nu^\prime}(t)
= \frac{1}{4\pi}\int_{\gamma_{\rm e,min}}^{\gM}
d\gamma_{\rm e} \,n_{\rm e}(\gamma_{\rm e},t) P^\prime_{\nu^\prime}(\gamma_{\rm e}).
\label{eq:j_nu}
\eeqn
Following \citet{hoeft07}, we introduce dimensionless variables, $\tau= C_{\tau} \gamma_{\rm e}$  and $\eta=\left(b_{\rm IC}/C_{\tau}\right) t$, where $C_{\tau}\equiv \sqrt{eB/(2\pi m_{\rm e}c \nu^{\prime})}\sim 4.3\times10^{-5}\ M_{\rm vir,13}^{1/3}\,\xi_{\rm B,-1}^{1/4}\,\nu_{9}^{-1/2}\,\hatrsh^{-9/20}\,\hatz^{-1/4}$ (so that $\nu/\nu_{\rm syn}=1/\tau^2$ and $b_{\rm IC} \gamma_{\rm e} t=\eta \tau$). Then, the integral in equation \ref{eq:j_nu} is proportional to
\beqn
J(\eta;s)\equiv\int_{\tau(\gamma_{\rm e,min})}^{1/\eta} d\tau \,\tau^{-s}(1-\eta\tau)^{s-2} F\left(\frac{1}{\tau^2}\right).
\eeqn
Note that $J(\eta;s)$ is a decreasing function of dimensionless time $\eta$ with $J(0;s)\sim0.8$--$1.0$ for $s=2.0$--$3.0$ (see Appendix \ref{s:J} for its detailed behaviors). Since $\tau(\gamma_{\rm e,min})=C_\tau \gamma_{\rm e,min}\ll1$ and $F(1/\tau^2)$  rapidly drops at $\tau\lesssim0.1$, we can effectively set $\tau(\gamma_{\rm e,min})=0$.

\begin{figure*}
\centering
\includegraphics[width = 17 cm, trim=5 19 5 5, clip]{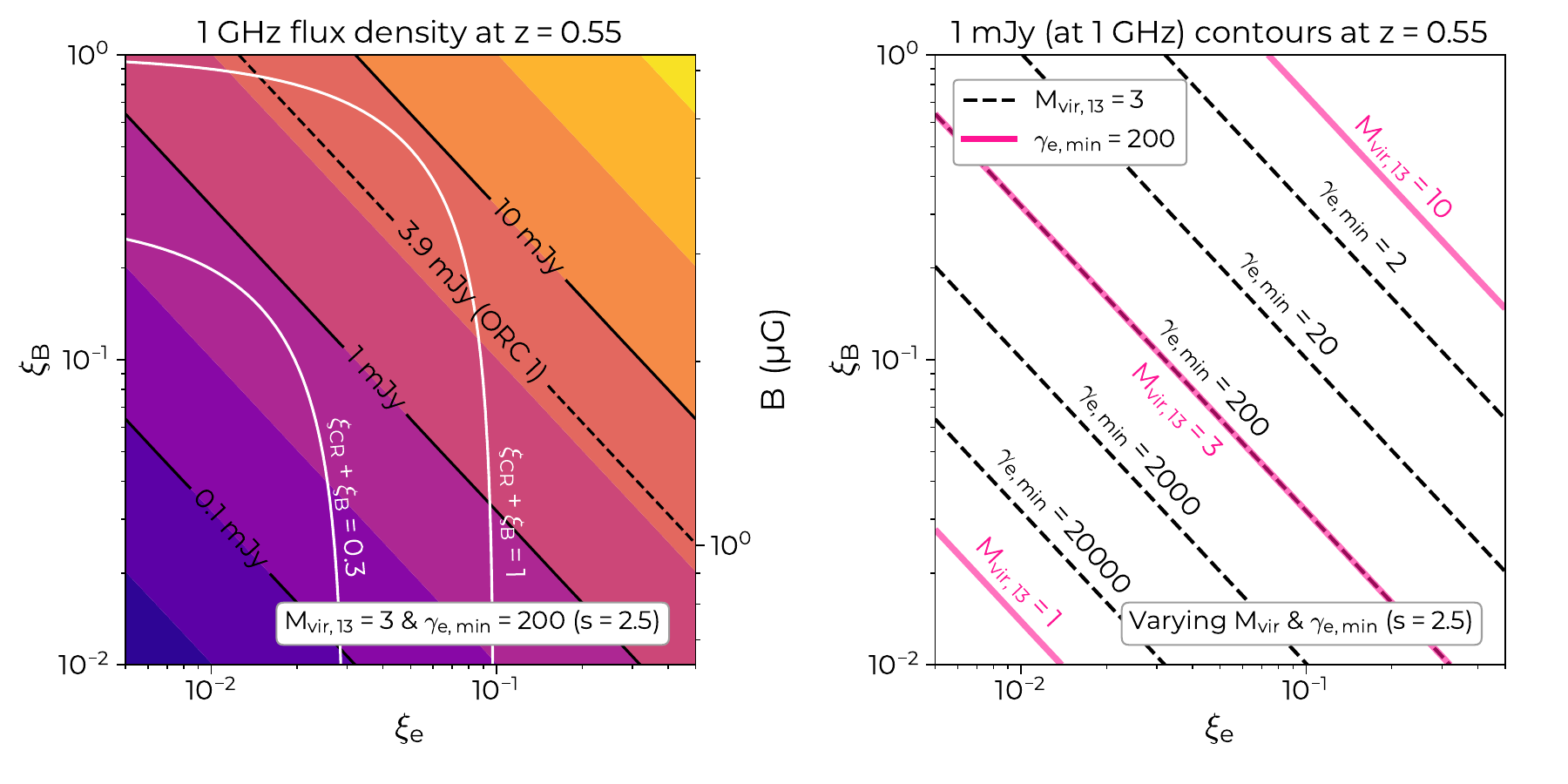}
\caption{{\it Left:} contours of flux density of a virial shock with fiducial parameters in ${\rm mJy}$ at observing frequency of $1$ GHz in the phase space of equipartition parameters for relativistic electrons, $\xi_e$, and for magnetic fields, $\xi_B$ with corresponding magnetic field strength in $\mu$G. 
The solid white curves represent conditions for different efficiency of energy conversion $\xi_{\rm CR}+\xi_{\rm B}$ (details in Appendix \ref{s:max_xi_e}).
{\it Right:} 1 GHz flux density contours plotted at $1$ mJy for different $M_{\rm vir}$ (magenta solid lines) and $\gamma_{\rm e,min}$ (black dashed lines).
The parameters that we fix or vary in calculating these quantities are indicated in each panel.
}
\label{fig:xi_plane}
\end{figure*}

Assuming that the downstream material is a steady flow (roughly), we can associate a given distance from the shock with a time, $t$, since the material passed through the shock. Consequently, the electron spectrum within a given volume element becomes independent of time. 
Considering this effect, we compute the specific intensity profile at a dimensionless impact parameter, $b\equiv r/r_{\rm sh}$, under the assumption of a spherical emitting shell geometry with an outer radius, $r_{\rm sh}$. Consider a line-of-sight (LoS) element $dl$, where $l$ is measured from the midpoint between the two points where the shock front ($r=r_{\rm sh}$) and the LoS intersect (see Fig.~\ref{fig:geometry}). The relation is given by $l=\sqrt{r(t)^2-(br_{\rm sh})^2}$, with the distance from the galactic center defined as
\beqn
\label{eq:advection length}
r(t)=r_{\rm sh}-v_{\rm d}t=r_{\rm sh}(1-\eta/\eta_{\rm adv}).
\eeqn
Here, the critical dimensionless time $\eta_{\rm adv}$ is defined at the shock crossing time of CR electrons, as $\eta_{\rm adv}=(b_{\rm IC}/C_\tau)(r_{\rm sh}/v_{\rm d})=(b_{\rm IC}/C_\tau)t_{\rm dyn}$.

We perform an integration of the specific emissivity along the LoS:
\beqn
\label{eq:profile}
I_{\nu} (b)&=&
\frac{1}{(1+z)^3}\int_{-r_{\rm sh}\sqrt{1-b^2}}^{r_{\rm sh}\sqrt{1-b^2}} dl j^{\prime}_{\nu^{\prime}}(t)\\
&=&
I_{\nu}^{\rm adv}\int_{0}^{\eta_{\rm adv}(1-b)}d\eta\frac{1-\eta/\eta_{\rm adv}}{\sqrt{(1-\eta/\eta_{\rm adv})^2-b^2}} J(\eta;s),\nonumber
\eeqn
where we utilize the relation $I^{\prime}_{\nu^{\prime}}=(1+z)^3I_{\nu}$ as $I_{\nu}/\nu^3$ is Lorentz invariant \citep[e.g.,][]{rybicki79} and introduce a numerical constant 
$I_{\nu}^{\rm adv}=2\,n_{\rm e0}/(4\pi)\, C_{\tau}^{s}\,C_{\rm syn}\,b_{\rm IC}^{-1}\, v_{\rm d}\,(1+z)^{-3}$.
Equation \ref{eq:profile} reduces to a surface brightness estimation using a basic one-dimensional slab geometry (e.g., Eq. 31 of \citealt{hoeft07}) at $b=0$, achieved by removing the factor of $2$ in $I_{\nu}^{\rm adv}$. Finally, we define the typical specific intensity of the shell by its full width at half maximum (FWHM) values of $I_\nu(b)$, which is then converted into the shell's flux density, $S_\nu$, by multiplying it with the solid angle subtended by the source, $\Delta\Omega=\pi(r_{\rm sh}/d_{A}(z))^2$, where $d_{A}(z)$ represents the angular diameter distance at redshift $z$.

The left panel of Figure \ref{fig:xi_plane} shows the flux density as a function of energy-equipartition parameters for a halo with fiducial parameters $M_{\rm vir}=10^{13.5}\Msun$, $\gamma_{\rm e,min}=200$, $r_{\rm sh}=260$ kpc, and $s=2.5$ located at $z=0.55$. The equipartition parameters for electrons $\xi_{\rm e}$, and for magnetic fields, $\xi_{\rm B}$, are not entirely independent, as the total sum of the cosmic-ray energy fraction, $\xi_{\rm cr}=\xi_{\rm e}+\xi_{\rm p}$ (where $\xi_{\rm p}$ is the proton energy density fraction, see Appendix \ref{s:max_xi_e} for the assumption for $\xi_{\rm p}/\xi_{\rm e}$), and $\xi_{\rm B}$ must not exceed 1. We show two cases: one representing maximum system efficiency, where $\xi_{\rm cr}+\xi_{\rm B}=1$, and another with reduced efficiency at 30\%, where $\xi_{\rm cr}+\xi_{\rm B}=0.3$ (as depicted in the left panel of Figure \ref{fig:xi_plane}). 
Considering these constraints, achieving an observed shell brightness of $\gtrsim1$ mJy requires values of $\xi_{\rm e}$ in the range of $0.01$ to $0.1$ and $\xi_{\rm B}\gtrsim0.1$ (corresponding to a magnetic field strength of $\gtrsim\mu$G). 

These constraints, of course, are also sensitive to the choice of other model parameters that cannot be easily inferred from the observations. For instance, the variation in phase space by choosing different values of $M_{\rm vir}$ and $\gm$ is shown in the right panel of Figure \ref{fig:xi_plane}. For a given $M_{\rm vir}$, a larger $\gm$ increases the number of required high-energy electrons, thereby lowering the energy conversion efficiency. For a given $\gm$, a more massive halo has more emitting electrons and hence loosens the energetic requirement, while a lower mass halo with $M_{\rm vir}\lesssim10^{12}\Msun$ is too faint to be detected regardless of $\gm$ due to the lack of enough electrons emitting in GHz band. Therefore, accretion shocks in less massive galaxies may not be visible in the radio band.
It is nonetheless true that the virial shocks of $M_{\rm vir} \gtrsim 10^{13} \Msun$ can produce radio emission that is consistent with the observations of ORCs.

\subsection{Cosmic-ray advection vs. diffusion}
\label{ss:shellwidth}

Apart from reproducing the size and intensity of the ORCs, one also has to reproduce the width of the radio rings as it might indicate important constraints. 
While the virial shock is typically stationary, the newly accelerated CR electrons can move by either advection with the shocked material behind the shock at a velocity of $v_{\rm d}=v_{\rm sh}/4\sim110\ {\rm km\ s^{-1}}\ M_{\rm vir,13}^{5/12}\, \hatrsh^{-1/4}\,h_{70}^{-1/4}\, \hatz^{1/8}$, or by diffusion along radially inward or outward. The actual value of the diffusion coefficient, $D_{\rm cr}$, for CRs is uncertain; while it is estimated to be around $\sim 10^{28}$ cm$^2$ s$^{-1}$ in the interstellar medium (ISM), it can be as high as $\sim 10^{29-31}$ cm$^2$ s$^{-1}$ in the low-density circumgalactic medium (CGM) \citep{Hopkins2021}. In the latter case, diffusion would dominate over advection. For simplicity, we assume that the diffusion coefficient is constant in both radial directions, i.e., the fractional shell width, $\delta = 2 \sqrt{4 D_{\rm cr} t_{\rm IC}}/r_{\rm sh}$, although, in reality, it may depend on the energy of the CR and the factor of $2$ could be a slight overestimation since the diffusion is happening against the flow in the upstream material.

\begin{figure}
\centering
\includegraphics[width=.48\textwidth, trim=0 7 0 0, clip]{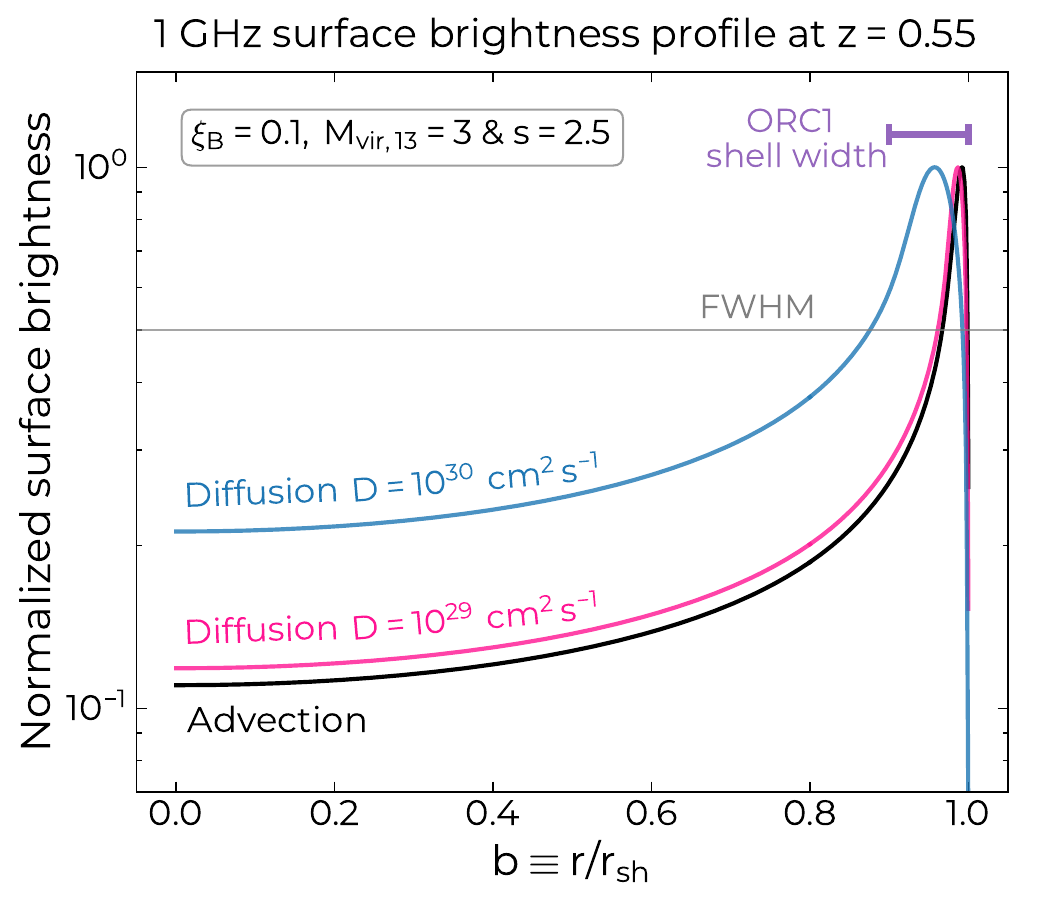}
\caption{$1$ GHz surface brightness profiles of virial shocks as a function of dimensionless impact parameter normalized by the peak brightness. The black curve represents the case of pure advection, computed based on equation \eqref{eq:profile}, while the red and blue curves correspond to diffusion cases, computed by a simple replacement of the advection length ($v_{\rm d} t$) with the diffusion length ($\sqrt{4D t}$) in equation \eqref{eq:advection length} and making appropriate modifications to equation \eqref{eq:profile} accordingly. The horizontal grey line represents the FWHM at which we define the average surface brightness.
The observed shell width of ORC1 is indicated in the top right corner of the plot. In our calculations, we maintain a fixed halo mass of $M_{\rm vir} = 3 \times 10^{13}\,{\rm \Msun}$, an electron power-law index of $s=2.5$, and a source redshift of $z=0.55$.
}
\label{fig:profile}
\end{figure}

In a pure advection scenario, we would expect the freshly shocked material to travel only up to a fractional width of the emitting shell, which can be expressed as:
\beqn
\delta&\approx& {v_{\rm d}\, t_{\rm IC}(\gamma_{\rm e,obs})\over r_{\rm sh}}\\&\sim& 0.009 \ M_{\rm vir,13}^{3/4}\,\xi_{\rm B,-1}^{1/4}\, \nu_{9}^{-1/2}\,\hatrsh^{-7/4}\,  h_{70}^{-1/4}\,\hatz^{-33/8}.\nonumber
\eeqn
Clearly, the shell width (FWHM) of the radio rings is at most $\sim 1\%$ of $r_{\rm sh}$, much less than the observed fractional shell width of $\sim 10\%$ \citep{norris22}\footnote{\citet{norris22} estimate that the intrinsic shell width for ORC1 after de-convolution of the radio beam is about 3--4 arcsec out of 40 arcsec, i.e. a fractional width of $9\%$.}. This suggests that the shell is not advection-dominated, but rather diffusion-dominated. 
Following this hypothesis, we can estimate the radio-emitting shell width is
\beqn
\delta&\approx& \frac{2\sqrt{4\, D_{\rm cr}\, t_{\rm IC}(\gamma_{\rm e,obs})}}{r_{\rm sh}}\\
&\sim& 7.9\times10^{-2} D_{{\rm cr},30}^{1/2}\,M_{\rm vir,13}^{1/6}\,\xi_{\rm B,-1}^{1/8}\,\nu_9^{-1/4}\,\hatrsh^{-1/4}\,h_{70}^{-1/4}\, \hatz^{-33/8}\ ,\nonumber
\eeqn
where $D_{{\rm cr},30}=D_{\rm cr}/(10^{30}\ {\rm cm^2\, s^{-1}})$.
Therefore, we require $D_{\rm cr} \approx 10^{30}$ cm$^2$ s$^{-1}$ to explain the observed shell width of ORCs (see Figure \ref{fig:profile} for numerical results).  Interestingly, the constraint on $D_{\rm cr}\propto M_{\rm vir}^{1/3}\,\xi_{\rm B}^{1/4}$ for a given value of $\delta$ only weakly depends on the model parameters $M_{\rm vir}$ and $\xi_{\rm B}$ (note that the shell width is independent of $\xi_{\rm e}$), suggesting that ORCs may provide a novel method to measure the diffusion coefficient in the CGM. Furthermore, within the thin layer considered here, the assumption of uniform magnetic fields (and hence a uniform diffusion coefficient) is a reasonable approximation.  

\section{Observable source number density}
\label{s:density}

\begin{figure}
\centering
\includegraphics[width=.49\textwidth, trim=0 15 0 0, clip]{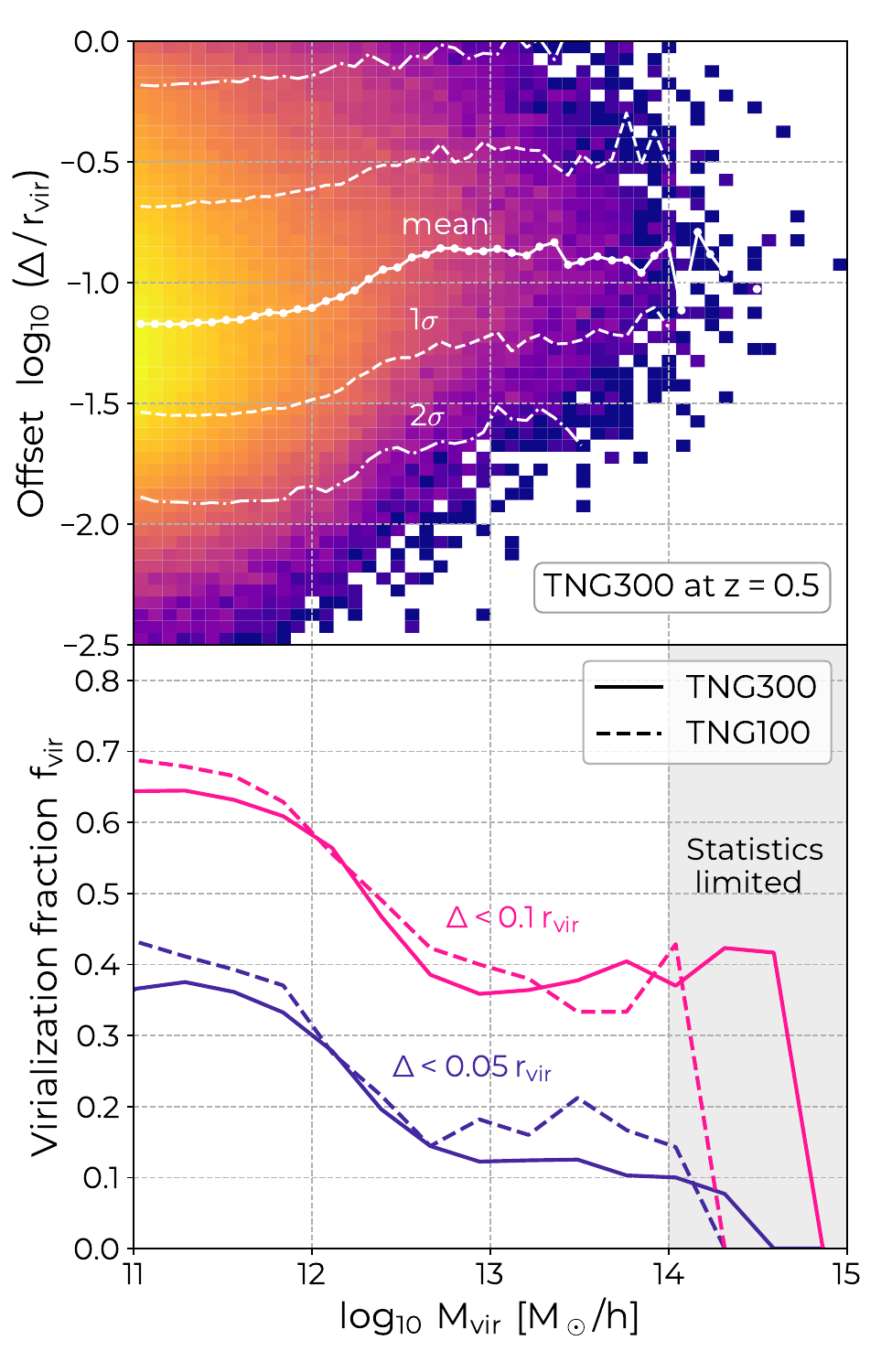}
\caption{{\it Top:} The offset $\Delta$ between the minimum potential point and the center of mass of a halo for TNG300 simulation at $z=0.5$. {\it Bottom:} The virialization fraction $f_{\rm vir}$ as a function of halo mass $M_{\rm vir}$. Since the definition of complete virialization (results for our choice $\Delta /r_{\rm vir}<0.05$ is shown in blue lines) is rather arbitrary, we also show the result for a looser condition $\Delta /r_{\rm vir}<0.1$ (red lines). The solid and dashed lines represent the TNG300 (low resolution) and TNG100 (high resolution) simulations, respectively. The grey-shaded regions are likely affected by the poor mass resolution (left) and statistics (right) of the simulation.  
}
\label{fig:f_vir}
\end{figure}

The observed comoving volume density of ORCs is roughly $1$ ORC per $0.05 \ {\rm Gpc^3}$, or $2\times10^{-8} \pcMpc$ assuming five ORCs are located at $z=0.3$--$0.6$ in the surveyed sky region \citep{norris22}. Since three of them are known to have central galaxies, we conservatively take the ORC number density at cosmological distances as $\Phi_{\rm ORC}\sim10^{-8}\pcMpc$. One of the most studied sources ORC1 has a massive central galaxy with a stellar mass of $M_\star \approx 3 \times 10^{11}\Msun$ at $z=0.55$ \citep{norris22}. 
A stellar-to-halo mass relation $M_\star/M_{\rm vir}$ \citep[e.g.,][]{moster10,wechsler2018,girelli20} implies that such a massive galaxy should possess a halo with a mass of $M_{\rm vir}\sim10^{13}\Msun$. The cumulative halo number density above $10^{13}\Msun$ is $\Phi_{\rm halo}\sim10^{-4}\pcMpc$ \citep{ps74,tinker2008}, 
which is four orders of magnitude higher than $\Phi_{\rm ORC}$. 
This apparent discrepancy requires some explanations why the ORCs from galactic virial shocks are not so common. 
\begin{itemize}

\item {\it Shock conditions? --} As shown in \S \ref{s:emission}, the observed radio flux density highly depends on the parameters that describe shock physics. In particular, a halo with a fiducial mass of $M_{\rm vir}\sim10^{13}\Msun$ must have a relatively high radiative efficiency (i.e., $\xi_{\rm e}$ and $\xi_{\rm B}$) of about $1$-$10$\%. A fraction of sources that achieve such high radiative efficiency under which virial shocks act as efficient accelerators of electrons could be relatively small. Namely, a slightly different set of $(\xi_{\rm e},\,\xi_{\rm B})$ compared to the values required by observations can readily make the source too faint to be detected, which might significantly reduce the number of observable ORCs. Considering this unknown factor for radiation efficiency $f_{\rm rad}$, the observable source number density could be significantly lower. We find that this factor is the primary reason for the underobservation of such ORCs.

\item {\it Incomplete virialization? --} 
It is possible that only perfectly virialized haloes can successfully form spherical virial shocks with associated emissions. We investigated the ``virilization fraction'' $f_{\rm vir}$ of haloes using TNG300/100 cosmological hydrodynamical simulation \citep{nelson19}, and assume that non-virialized haloes may not create spherical accretion shocks but rather one multiple radio arcs. We use the offset $\Delta$ between the minimum potential point and the center of mass of a halo as a rough proxy for the level of virialization (a halo with smaller $\Delta$ is better virialized). Here, we define haloes with $\Delta <0.05 \,r_{\rm vir}$
as virialized
(cf.\ e.g., \citealt{neto2007,cui2018}). Figure \ref{fig:f_vir} shows the virialization fraction of TNG haloes at $z=0.5$ as a function of halo mass.
We obtained $f_{\rm vir}\sim0.15$ 
for  
a criterion $\Delta<0.05 \,r_{\rm vir}$ 
for haloes with $M_{\rm vir}=10^{13}\,\Msun$ at $z=0.5$.
Note that the mass range above $10^{14}\Msun$ is likely suffering from poor statistics. Nevertheless, we see that $f_{\rm vir}$ generally decreases as $M_{\rm vir}$ increases. 
Also, for a given halo mass, $f_{\rm vir}$ is generally smaller at higher $z$ due to the more active halo growth and frequent merger events. 

Note that the above estimate of spherical virial shocks is a preliminary approximation.  It remains unclear if $\Delta$ is a good indicator for spherical virial shocks or what
threshold should be adopted, which is worth future investigation
with pertinent hydrodynamical simulations. 
Radio relics, particularly ORC-like relics, are relatively rare in the broader context. Observations and simulations suggest that radio relics constitute a minor fraction (approximately $\lesssim$10\% for $10^{14}\Msun$ halos; \citealt{Kale2015, Jones2023, Lee2023}), likely even less so in lower mass halos. 
Furthermore, galaxy clusters exhibit a diverse range of radio relic morphologies and properties, including double and single relics, linear and arc-shaped systems, complicating the overall shock geometry, especially in the presence of mergers and filaments \citep{Lee2023}. The radii of shocks can vary based on direction \citep[e.g.,][]{Molnar2009, Power2020}, and the observed morphology is dependent on the viewing angle. During mergers, merger-accelerated accretion shocks form and rapidly propagate to a few virial radii, where the density is lower and, consequently, less observable. Genuine accretion shocks are present only during relatively quiescent periods of halo evolution \citep{Zhang2020}, which motivates us to seek relaxed halos.
Presently, there is no precise estimate of the frequency of ORCs based on current simulations. Several factors contribute to this uncertainty: (1) Simulation results are often derived from a limited sample of halos or involve stacked samples without detailed statistics on different morphological types. (2) Compared to galaxy clusters ($>10^{14}\Msun$), there is much less attention paid to the accretion shocks of $10^{13}\Msun$ halos. (3) The current simulations are subject to model uncertainties, e.g., in the treatment of cosmic rays and feedback models \citep[e.g.,][]{Ji2021,Power2020}.

\end{itemize}

Taking all the above into consideration, the number density of observable circular radio emission from virial shocks might be estimated by $\Phi_{\rm vir}\approx f_{\rm rad}\,f_{\rm vir}\, \Phi_{\rm halo}(>M_{\rm vir})$, where $\Phi_{\rm halo}(>M_{\rm vir})$ is the cumulative halo mass function above a given halo mass $M_{\rm vir}$.
With a fiducial halo mass of $M_{\rm vir}\sim10^{13}\Msun$ this yields 
\beqn
\Phi_{\rm vir}\sim10^{-6}\pcMpc \left(\frac{f_{\rm rad}}{0.1}\right)\, \left(\frac{f_{\rm vir}}{0.1}\right)\left(\frac{\Phi_{\rm halo}}{10^{-4}\pcMpc}\right).
\eeqn
Given the model uncertainties in $\Phi_{\rm vir}$ arising from unknown $f_{\rm rad}$ ($<1$) and $\Phi_{\rm ORC}$ estimated from only several events, virial shocks could be a viable option to accommodate the observed number density of ORCs (i.e., $\Phi_{\rm ORC}\ll\Phi_{\rm vir}$), but the fraction of virial shocks producing observable ORCs ($f_{\rm rad}f_{\rm vir}$) must be of much smaller than unity depending on the halo mass. We will need a better understanding of particle acceleration and their lifetime to refine the estimates.

The number density of ORCs should also have a strong dependence on halo mass. Although high mass haloes ($M_{\rm vir}\gg10^{13}\Msun$) are more capable of producing bright emission, the number density dramatically decreases down to $10^{-8}\pcMpc$ at $M_{\rm vir}\sim10^{15}\Msun$ \citep[e.g.,][]{ps74,tinker2008} and their virialization fraction becomes lower ($f_{\rm vir}\ll1$), which makes the number density of observable population much smaller 
than that for $10^{13}\Msun$ haloes. Meanwhile, lower mass halos ($M_{\rm vir}\ll10^{13}\Msun$) suffer from the lack of enough electrons to produce radio emission, which cannot overcome the abundance of halo and thus result in the decrease in the observable number density. Additionally, there is a cutoff in halo mass at a few $10^{11}\Msun$, below which spherical virial shocks do not form 
\citep{Birnboim2003,keres05}. These may be why we particularly observe ORCs in $10^{13}\Msun$ haloes with relatively high radiative efficiency.

\section{Discussion}
\label{s:discussion}

\subsection{Comparison with \citet{dolag22}}

While finalizing this manuscript, \citet{dolag22} proposed a possibility of merger-driven internal shocks as an origin of ORC structures with $M_{\rm vir}=10^{12}\,\Msun$ situated at $z\sim0.5$. While it accounts for both the rarity of ORCs and the complex inner sub-structure seen in ORCs, their numerical emission model fails to explain the observed synchrotron emission brightness. According to their estimation, the GHz power of the ring is approximately $10^{12}$--$10^{13}$ W$/$Hz. Assuming a very rough shell area of $A\sim (300 {\ \rm kpc})^2 \sim 8\times10^{47}\ {\rm cm^2}$ based on their simulation (see Fig. 8 in \citealt{dolag22}), the resulting surface brightness $S\sim P/A$ falls in the range of $1$--$10\ {\mu}$Jy. This range is only $2$--$3$ orders of magnitude fainter than the observed emission ($\sim$mJy). 

They assumed electron LF in the range of $10^2<\gamma_{\rm e}<10^5$, which aligns with our assumptions. They also considered a constant magnetic field of $5\,\mu$G, which is five times higher than ours and close to the CMB magnetic field $B_{\rm CMB}\sim7.2\,\hatz^2\,\mu$G. Additionally, they adopted an electron-to-proton number density ratio ($\xi_{\rm e}/\xi_{\rm p}=10^{-2}$ (in our case, this ratio is approximately $0.1$, as detailed in Appendix \ref{s:max_xi_e}). The higher magnetic field in their model increases the synchrotron power, while the roughly one order of magnitude smaller efficiency for accelerating relativistic electrons would reduce the number of electrons. These effects somewhat cancel each other out in terms of the radio ring brightness. 

Most importantly, they consider a $10^{12} \Msun$ halo, which is less massive than the $10^{13} \Msun$ halo in our model\footnote{\citet{dolag22} opted for a less massive halo, resulting in a virial radius $r_{\rm vir}\sim 200$ kpc $M_{\rm vir,12}^{1/3}\,\hatz^{-1/2}$, which aligns with the ORC ring radius. In contrast, our model yields a virial radius $r_{\rm vir}\sim 400$ kpc $M_{\rm vir,13}^{1/3}\,\hatz^{-1/2}$, potentially twice as large as the observed value. However, we consider the actual shock radius $r_{\rm sh}$ (ORC radius) to be potentially much smaller than the virial radius (see \S \ref{subsec:virial-shocks}).}.
Their consideration of a smaller halo mass reduces the thermal energy of the shock, which thereby reduces the CR energy density and, therefore, results in much fainter synchrotron emission (see also right panel of Figure \ref{fig:xi_plane}) compared to what is observed (or what is predicted by our model). In summary, their results could potentially remain consistent with ours if certain parameters are tuned.
We, therefore, encourage the galaxy merger search conducted in \citet{dolag22} should also be repeated for haloes with a mass of $10^{13} \Msun$.

\subsection{Hadronic Emission?}
Although we mainly discussed primary relativistic electrons produced at the shock, the interaction of shock-generated protons with cold ones behind the shock can also keep supplying secondary relativistic electrons. 
Injected protons cool via inelastic \textit{pp} collisions with cold protons in the post-shock region (with number density $n\approx 4\rho_{\rm sh}/m_{\rm p}$) at a rate
$\dot{\gamma}_{\rm p}=-\left(0.5\,\sigma_{\rm pp} \,n\, c \right)\,\gamma_{\rm p}$, 
where we assumed a constant elasticity $0.5$ and neglected the weak energy dependence of the cross section $\sigma_{\rm pp}\sim3\times10^{-26}\ {\rm cm^2}$ \citep{mannheim94}. The cooling time of protons is thus energy independent and estimated as $t_{\rm pp}=\gamma_{\rm p}/|\dot{\gamma}_{\rm p}|\sim73 \ {\rm Gyr}\ \,M_{\rm vir,13}^{-1/2}\,\hatrsh^{3/2}\,h_{70}^{3/2}\, \hatz^{-3/4}$. Hence, the number of such electrons is smaller roughly by a factor of $t_{\rm IC}(\gamma_{\rm e,obs})/t_{\rm pp}\sim{\cal O}(10^{-3})$ than the primary populations, thereby resulting in a negligible contribution to the surface brightness of the ring.

\subsection{Polarization}
The linear polarization fraction of an optically thin source emitting synchrotron
radiation in a uniform field is at most $(s+1)/(s+ 5/3)\sim0.69$--$0.77$ for $2<s<3$. However, the observed mean polarization fraction for ORC1 is approximately $30$\% \citep{norris22}, considerably below the theoretical maximum. Similar deviations from the theoretical maximum have been noted in radio polarization studies of supernova remnants, where comparable particle acceleration and magnetic amplification processes are anticipated \citep{reynoso13}. A plausible explanation for this deviation is beam depolarization, an instrumental effect in which polarized vectors within a finite beam size counteract one another, resulting in a decrease in the observed polarization fraction after convolution with the telescope beam \citep{sokoloff1998}. Even
though the magnetic fields appear ordered at the level of MeerKAT image resolution, the low fractional polarization hints at a more general disorganization of magnetic fields on smaller scales within these regions. If there exists a turbulent magnetic field component with strength $B_{\rm turb}$ in addition to the ordered $B$, this would lead to a reduction in the polarization fraction by a factor of approximately $\sim B_{\rm turb}^2/(B^2+B_{\rm turb}^2)$.  
Observing ORCs at sufficiently high angular resolutions can mitigate the impact of beam depolarization and yield stringent constraints on the orientation of the magnetic field and the magnitude of $B_{\rm turb}$.

\subsection{Pre-Heating Electrons?}

Intriguingly, our result implies that the low-energy cutoff energy for relativistic electrons at injection could be relatively large $\gamma_{\rm e,min}\gtrsim100$. Recent particle-in-cell simulation results \citep{xu20,kumar21} show evidence for the early onset of the diffusive shock acceleration (DSA) process for both electrons and ions at oblique non-relativistic shocks. \citet{xu20} show that electrons undergo multiple cycles of shock-drift acceleration due to their reflection by downstream magnetic fields and the subsequent growth of upstream turbulence, which traps and pre-heats the electrons before the injection into DSA. Therefore, our results suggest that ORCs might highlight the less-understood role of pre-heating electrons in non-relativistic shocks.

\subsection{Caveats}
Despite the overall consistency with ORCs, there are a few potential caveats in our model. 
The main caveat is that we did not solve CR transport in a spherical geometry.
Also, a consideration of the CR transport that includes both advection and diffusion simultaneously is important for the realistic calculation of the brightness profile, which we defer for future work.
The second is the assumption of correlation between gas energy density and CR energy density. Since our brightness estimate relies on the CR energy density, which we assumed to be proportional to the gas energy density at the virial shock, more elaborate modeling of CR acceleration at the virial shock would be needed. However, we speculate that this would change our estimate only by some factor.

\section{Summary \& Conclusion}
In this work, we present a new model in which cosmological ORCs are produced by the synchrotron emission from leptonic virial rings around massive galaxies. Our model is described by several parameters, the halo mass $M_{\rm vir}$, the shock radius $r_{\rm sh}$, the power-law index of electron energy distribution $s$, the minimum electron LF $\gm$, and the fraction of shock energy transferred to relativistic electrons $\xi_{\rm e}$ and to downstream magnetic field $\xi_{\rm B}$. The electron index could be inferred from the radio observation (although the error is large), thus we are left with four.

Our findings are summarized as follows:

\begin{itemize}
    \item We show that observed size ($200$ kpc radius) and brightness ($\sim$mJy at GHz) of ORCs are broadly consistent with virial shocks with $M_{\rm vir}\sim10^{13}\Msun$ with $s\sim2$--$3$.
    
    \item We also find that pure advection of electrons from the shock results in a radio-emitting shell that is considerably narrower than the observed one due to strong inverse-Compton cooling of electrons. Instead, we show that the diffusion of CR electrons plays a significant role in increasing the width of the shell. We infer a diffusion coefficient, $D_{\rm cr} \sim 10^{30}\ {\rm cm^2\,s^{-1}}$, consistent with the values expected for low-density CGM. If the ORCs indeed originate from the virial shocks, then our estimated CR diffusion coefficient is one of the very few estimations obtained for the low-density CGM. 
    
    \item Furthermore, we show that the apparent discrepancy between ORC ($\sim10^{-8}\pcMpc$) and galaxy (or halo) number density $\sim10^{-4}\pcMpc$ for $10^{11}\Msun$ galaxies (or for $10^{13}\Msun$ haloes) can be mitigated by considering both the incomplete virialization of haloes as suggested by cosmological simulations and the limited radiation efficiency of shocks. 
    
    \item We show that low-mass galaxies ($\lesssim 10^{13}\Msun$) may not appear as ORCs due to their very low radio surface brightness. This may be why we particularly observe ORCs in $10^{13}\Msun$ haloes with relatively high radiative efficiency. 

\end{itemize}

In conclusion, we presented a simple but generic model for synchrotron virial shocks and showed that it can broadly account for the observed ORC properties. We found that ORCs could offer a promising avenue for investigating CR diffusion in low-density CGM. Our qualitative model predictions could be tested with an expanded sample of ORCs, especially with the advent of Square Kilometer Arrays (SKA).
In future investigations, the incorporation of both advection and diffusion in numerical simulations of CR transport in spherical geometry holds the potential to yield more realistic calculations of the brightness profile.
Moreover, we encourage spectroscopic observations targeting galaxies located within ORC rings in the south hemisphere. These observations are critical for accurately determining the redshift and group mass of central galaxies, shedding further light on the ORC-galaxy connection. Furthermore, we encourage continued exploration of galaxy mergers within simulations, particularly those capable of successfully replicating the circular ORC morphology, as conducted by \citet{dolag22}, in haloes with masses around $\sim 10^{13} \Msun$.

\section*{Acknowledgements}
We express our gratitude to Matthias Hoeft and Marcus Brüggen for engaging in discussions regarding the radio-emitting shell width in their model. Special thanks to Bärbel Koribalski for providing information on the spectral index measurements of ORC1, and to Norita Kawanaka for their valuable comments during the early draft stage. SY acknowledges the support of Yu-An Chen in investigating ORC host galaxy candidates, Alvina On for initiating discussions on ORC polarization fraction, and Yen-Hsing Lin, Hsiang-Yi Karen Yang, and Shutaro Ueda for discussions at the ASROC Annual Meeting and NCHU. SY also appreciates Tomotsugu Goto, Tetsuya Hashimoto, and Tomonori Totani for early discussions and encouragement. Finally, we thank the referee for careful reading and useful suggestions.
SY acknowledges support from the National Science and Technology Council of Taiwan through grants 110-2112-M-005-013-MY3, 110-2112-M-007-034-, and 112-2123-M-001-004-. KCS is supported by the German Science Foundation via DFG/DIP grant STE/ 1869-2 GE/ 625 17-1 in Israel.
ZL is supported by the Israel Science Foundation Grant ISF 861/20 and the European Union’s Horizon 2020 research and innovation programme under the Marie Skłodowska-Curie grant agreement No 101109759 (``CuspCore'').


\section*{Data Availability}
The TNG300/100 cosmological hydrodynamical simulation data are available from \url{https://www.tng-project.org/}. Custom code will be made available upon reasonable request to the corresponding author.



\bibliographystyle{mnras}
\input{orc.bbl}



\appendix

\section{Scaling relations}

Below we show some useful scaling relations for density profiles employed in this study. To determine the dark matter halo profile, we adopt mass-concentration relations from \citep{zhao09,dutton14} and consider a concentration parameter $c$ ranging from 4 to 10 for halos with $M_{\rm vir}\lesssim10^{13}\,\Msun$ at redshifts $z\lesssim1$.

\subsection{NFW Profile}
\label{s:scalings}
The NFW halo density profile and enclosed mass profile respectively are given by
\beqn
\rho_{\rm dm}(r)&=&\frac{M_{\rm vir}}{4\pi A(c)}\frac{1}{r(r+r_{\rm s})^2},\\
M_{\rm dm}(<r)&=&
\frac{M_{\rm vir}}{A(c)}\left[\ln \left(1+\frac{r}{r_{\rm s}}\right)-\frac{r/r_{\rm s}}{1+r/r_{\rm s}}\right],
\eeqn
where $A(c)=\ln (1+c)-c/(1+c)$ and $c\equiv r_{\rm vir}/r_{\rm s}$ is the concentration parameter with $r_{\rm s}$ being the NFW scale radius. 
We find that the approximations $\rho_{\rm dm}(r)\propto r^{-2.6}$ and $M_{\rm dm}(<r)\propto \int  r^2\,\rho_{\rm dm}(r) dr\propto r^{0.7}$ hold for concentration parameters of $c=4$--$10$ at the shock radius (which is smaller than virial radius as defined in \S \ref{subsec:virial-shocks}) of $0.5<r_{\rm sh}/r_{\rm vir}<1$ to within $6$\% and $10$\% accuracy, respectively.

\subsection{Dark matter density at virial shock}
\label{s:rho_dm}
Comparing actual density at virial radius with the mean density inside it, i.e., $\overline{\rho}_{\rm dm}=M_{\rm vir}/(4/3\pi r_{\rm vir}^3)=200\rho_{\rm c}$, one gets
\beqn
\frac{\rho_{\rm dm}(r_{\rm vir})}{\overline{\rho}_{\rm dm}}=\frac{1}{3 A(c)}\left(\frac{c}{1+c}\right)^2.
\eeqn
Typically $\rho_{\rm dm}(r_{\rm vir})/\overline{\rho}_{\rm dm}\sim0.26$--$0.19$ for $c=4$--$10$, which means the ratio barely depends on the halo properties. Thus, we approximate the density at the virial radius by $\rho_{\rm dm}(r_{\rm vir})\sim0.25\, \overline{\rho}_{\rm dm}$. With a power-law dependence of $\rho_{\rm dm}(r)\propto r^{-2.6}$, the dark matter density at the shock radius, $r_{\rm sh}$, might be estimated as $\rho_{\rm dm}(r_{\rm sh})= \rho_{\rm dm}(r_{\rm vir}) (r_{\rm sh}/r_{\rm vir})^{-2.6}\sim0.25\,\overline{\rho}_{\rm dm} \,(r_{\rm sh}/r_{\rm vir})^{-2.6}$.

\subsection{Gas density at virial shock}
\label{s:rho_sh}

We assume that the circumgalactic medium (CGM) is distributed as adiabatic gas with polytropic index of $5/3$ that is in hydrostatic equilibrium within the NFW dark matter halo of concentration $c$ \citep{mb2004}:
\beqn
\rho_{\rm gas}(r)\propto\left[1+\frac{3.7}{r/r_{\rm s}}\ln \left(1+\frac{r}{r_{\rm s}}\right)-\frac{3.7}{c}\ln \left(1+c\right)\right]^{3/2}.
\eeqn
We normalize the profile such that $\int_{0}^{r_{\rm sh}} 4\pi r^{2}\,\rho_{\rm gas}(r) dr=f_{\rm b} \,M_{\rm dm}(<r_{\rm sh})$ without losing generality. In contrast to NFW-like gas halo, the adiabatic gas density distribution is flatter: $\rho_{\rm gas}(r_{\rm sh})=B(r_{\rm sh}; c)\,f_{\rm b}\,\overline{\rho}_{\rm dm}$, where $B(r_{\rm sh};c)\sim0.7\,(r_{\rm sh}/r_{\rm vir})^{-1.5}$ for $c=4$--$10$ at large shock radii $0.5<r_{\rm sh}/r_{\rm vir}<1$. We assume that this hydrostatic CGM profile $\rho_{\rm gas}(r)$ approximately holds till $r<r_{\rm sh}$ and regard it as the ``shocked'' matter density. Therefore, the shocked (downstream) gas density near the shock, $\rho_{\rm sh}$, is  $\rho_{\rm sh}=\rho_{\rm gas}(r_{\rm sh})\sim 0.7\,f_{\rm b}\,\overline{\rho}_{\rm dm} (r_{\rm sh}/r_{\rm vir})^{-3/2}$.  Then, the unshocked (upstream) matter density just above the shock is given by $\rho_{\rm sh}/4$.

\section{Synchrotron emissivity function}
\label{s:emissivity}
\begin{figure}
\centering
\includegraphics[width=.5\textwidth, trim=0 5 0 0, clip]{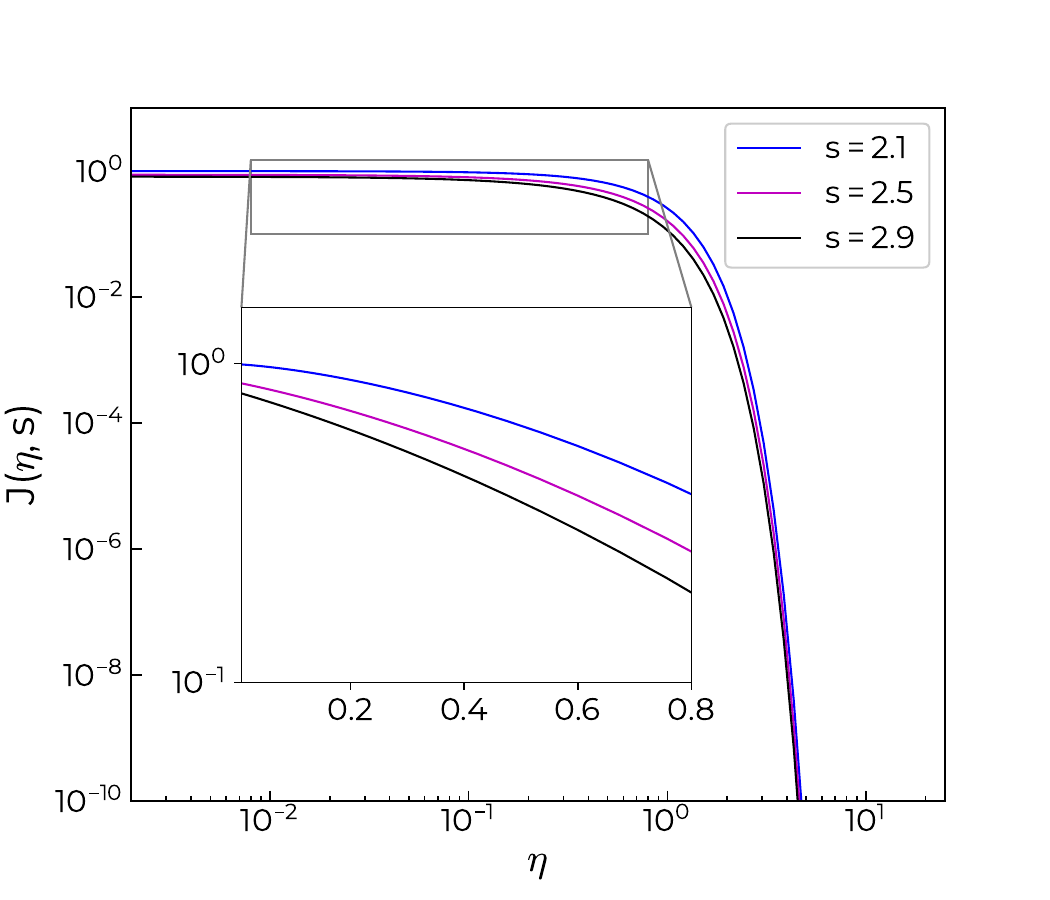}
\caption{$J(\eta;s)$ as a function of $\eta$ for different values of $s$.
}
\label{fig:J_eta}
\end{figure}

The synchrotron emissivity function is defined as
\label{s:J}
\beqn
J(\eta;s)\equiv
\int_{\tau_{\rm min}}^{\tau_{\rm max}} d\tau \,N(\tau;\eta,s) F\left(\frac{1}{\tau^2}\right),
\label{eq:J_nu}
\eeqn
where
\beqn
N(\tau;\eta,s)\equiv
\begin{cases}
\tau^{-s}(1-\eta\tau)^{s-2} & (\tau<1/\eta),\\
0& ({\rm otherwise})
\end{cases}
\label{eq:N}
\eeqn
is nearly equivalent to $n_{\rm e}(\gamma_{\rm e},t)$ in equation \ref{eq:n_e_primary}. As shown in \S \ref{s:emission}, both integration limits can be regarded as $\tau_{\rm min}=0$ and $\tau_{\rm max}=\infty$. In numerical computation, the integration is performed with log-uniform bins spanning over $\tau\in[\tau_{\rm min},\tau_{\rm max}]$ with $\tau_{\rm min}=10^{-3}$ and $\tau_{\rm max}=10^{4}$. The lower integration limit, $\tau_{\rm min}=C_\tau\gamma_{\rm e,min}$, is chosen so that it is always below  $1/\eta\geq1/\eta_{\rm c}\sim3.7\times10^{-3}\ \nu_{9}^{-1/2}\,f_{\rm sh}^{-7/4}\ M_{\rm vir,13}^{1/6}\,\xi_{\rm B,-1}^{1/4}\,h_{70}^{3/2}\,\hatz^{-13/4}$. The result is insensitive to the choice of $\tau_{\rm min}$ as long as $\tau_{\rm min}\lesssim0.1$ since $F(x)$ drops very fast at $x\gtrsim100$.   
As shown in Figure \ref{fig:J_eta}, $J(\eta;s)$ is a decreasing function of dimensionless time $\eta$ and $J(0;s)\sim0.8$--$1.0$ for $s=2.0$--$3.0$.

\section{Proton-to-electron energy density ratio}
\label{s:max_xi_e}
\citet{persic15} determines the energy density ratio between protons and electrons as $\xi_{\rm p}/\xi_{\rm e} = (m_{\rm p}/m_{\rm e})^{(3-s)/2}$, assuming that they share the same power-law index above $10$ keV and that the charge neutrality holds. This, combined with trivial limit $\xi_{\rm e}+\xi_{\rm p}+\xi_{\rm B}<1$, sets an upper-limit on electron energy fraction as $\xi_{\rm e}\le1/\left[1+(m_{\rm p}/m_{\rm e})^{(3-s)/2}\right]$.
Since our assumption on the electron energy distribution is different from theirs, here we derive the appropriate ratio for our case. Let us assume that the protons are injected with the same index as electrons above a minimum LF $\gamma_{\rm p,min}$, i.e.,  $n_{\rm p}(\gamma_{\rm p})= n_{\rm p0}\, \gamma_{\rm p}^{-s}$.
The charge neutrality condition reads
\beqn
    \int_{\gamma_{\rm e,min}}^{\infty} n_{\rm e}(\gamma_{\rm e}) d\gamma_{\rm e}=\int_{\gamma_{\rm p,min}}^{\infty} n_{\rm p}(\gamma_{\rm p}) d\gamma_{\rm p}.
\eeqn
This implies $n_{\rm p0}/n_{\rm e0}=(\gamma_{\rm p,min}/\gamma_{\rm e,min})^{s-1}$. Then, the energy density ratio between protons and electrons is
\beqn
    \frac{\xi_{\rm p}}{\xi_{\rm e}}&=&\frac{\int_{\gamma_{\rm p,min}}^{\infty} \gamma_{\rm p}m_{\rm p}c^2\,n_{\rm p}(\gamma_{\rm p}) d\gamma_{\rm p}}{\int_{\gamma_{\rm e,min}}^{\infty}\gamma_{\rm e}m_{\rm e}c^2\, n_{\rm e}(\gamma_{\rm e}) d\gamma_{\rm e}}=\frac{n_{\rm p0} m_{\rm p} \gamma_{\rm p,min}^{-s+2}}{n_{\rm e0} m_{\rm e} \gamma_{\rm e,min}^{-s+2}}\nonumber \\ 
    &=&\frac{\gamma_{\rm p,min} m_{\rm p}}{\gamma_{\rm e,min} m_{\rm e}}\sim 1.8 \left(\frac{\gamma_{\rm p,min}}{1}\right)\left(\frac{\gamma_{\rm e,min}}{10^3}\right)^{-1}.
\eeqn
The difference between \citet{persic15} is that they considered the thermal population of electrons but we do not (because this significantly reduces the number of available high-energy electrons that explains GHz observation). 

\bsp	
\label{lastpage}
\end{document}